\documentclass[fleqn,usenatbib]{mnras}

\usepackage{newtxtext,newtxmath}

\usepackage[T1]{fontenc}

\DeclareRobustCommand{\VAN}[3]{#2}
\let\VANthebibliography\thebibliography
\def\thebibliography{\DeclareRobustCommand{\VAN}[3]{##3}\VANthebibliography}

\usepackage{graphicx}	
\usepackage{amsmath}	
\usepackage[dvipsnames]{xcolor}

\title[Supernova connection of LHAASO J2108+5157]{Supernova connection of unidentified ultra high energy gamma-ray source LHAASO J2108+5157}

\author[A. De Sarkar]{
Agnibha De Sarkar,$^{1}$\thanks{E-mail: agnibha@rri.res.in}
\\
$^{1}$Astronomy $\&$ Astrophysics group, Raman Research Institute \\
C. V. Raman Avenue, 5th Cross Road, Sadashivanagar, Bengaluru 560080, Karnataka, India\\
}

\date{Accepted XXX. Received YYY; in original form ZZZ}

\pubyear{2022}

\begin{document}
\label{firstpage}
\pagerange{\pageref{firstpage}--\pageref{lastpage}}
\maketitle

\begin{abstract}
We present a simple phenomenological model of hadronic interaction between protons accelerated in an old supernova remnant (SNR) and cold protons situated within the associated molecular clouds (MCs). The accelerated protons from the old SNR escaped the SNR shock front, and got injected into the MCs at an earlier time, producing ultra high energy gamma-rays and neutrinos through inelastic proton-proton interaction. We also take into account the acceleration and subsequent escape of electrons from the SNR shock front. The escaped electrons produce gamma-rays through various radiative cooling mechanisms, after getting injected into the MCs. We use the model discussed in this letter to explain the multiwavelength (MWL) spectral energy distribution (SED) of unidentified Galactic ultra high energy gamma-ray source LHAASO J2108+5157. We also discuss the feasibility of applying this model in other cases as well. Future observations can test the viability of the model discussed in this letter, which will in turn confirm that the SNRs can, in fact, accelerate particles up to PeV energies.
\end{abstract}




\section{Introduction}\label{sec1}

Observations by the Large High Altitude Air Shower Observatory (LHAASO), located in China, have opened a new era of gamma-ray astrophysics \citep{lhaaso10}. Since it has become operational on 2020 April, LHAASO has detected more than a dozen ultra high energy (UHE; E$_\gamma$ $\geq$ 100 TeV) gamma-ray sources, most of which are unidentified \citep{Cao21}. The detection of these UHE gamma-ray sources indicates the presence of cosmic ray (CR) accelerators in our Milky Way Galaxy, which can accelerate particles up to PeV (= 10$^{15}$ eV) energies, more commonly known as ``PeVatrons''. Several classes of Galactic sources such as supernova remnants (SNRs), pulsar wind nebulae (PWNe), young stellar clusters have been posited to be potential PeVatron candidates. Although it is still an open question as to what class of source is responsible for accelerating particles up to PeV energies, most of the UHE gamma-ray sources detected by LHAASO, along with their high energy (HE; E$_{\gamma}$ $<$ 100 GeV) and very high energy (VHE; 100 GeV $\leq$ E$_{\gamma}$ $<$ 100 TeV) gamma-ray counterparts, have been associated with PWNe in previous studies, due to their close proximity with an energetic pulsar, and their typically extended spatial morphology \citep{hess18}. This idea gained steam after it was confirmed that Crab pulsar wind nebula is indeed a PeVatron source \citep{Cao21}. However, in spite of the notion that energetic pulsars with high spin-down luminosity ($\dot{E}$ $>$ 10$^{36}$ erg s$^{-1}$) coinciding or remaining in a very close spatial proximity of UHE gamma-ray sources, may be a universal feature \citep{albert21}, further theoretical analyses of these LHAASO detected UHE gamma-ray sources seem to tell a different story.   

Recent studies have modeled a few of the significantly detected LHAASO sources in detail. For example, \cite{desarkar22a} found that the UHE gamma-ray emission observed from LHAASO J1908+0621 is most likely hadronic in origin, emanated from the interaction between SNR G40.5-0.5 and the associated MCs. On the other hand, in \cite{desarkar22b}, another significantly detected source, LHAASO J2226+6057, was extensively modeled assuming that the UHE gamma-ray emission is coming from the PWN associated with PSR J2229+6114. As caveats of the model, it was found that the PWN interpretation of LHAASO J2226+6057 leads to a very high radius of PWN, as well as a very small value of magnetic field. Naturally, these results are in contrast with the observational results \citep{ge21, liang22}, thus indicating that the PWN may not be the contributing source to power the UHE gamma-ray source detected. This indicates interaction between SNRs and associated MCs may be the primary reason behind particle acceleration to PeV energies in Galactic sources. 

With this factors in mind, we focus on the emission of the recent LHAASO detected unidentified UHE gamma-ray sources: LHAASO J2108+5157 \citep{cao2108} and LHAASO J0341+5258 \citep{cao0341}. Both of these sources were found to be associated with MCs, but no apparent association with energetic pulsars or SNRs were established. Scenarios including leptonic emission from TeV halo \citep{abe22}, injection of particles from past explosions \citep{kar22}, hadronic interaction between SNR and MCs \citep{cao2108} were discussed in previous literatures. But most of these models do not explain the HE-VHE-UHE gamma-ray spectral features entirely.  Moreover, recent reveal of VHE gamma-ray upper limits observed by the Large-Sized Telescope - Cherenkov Telescope Array (LST-CTA) \citep{abe22} has overruled some of these models for the case of LHAASO J2108+5157. The absence of a powerful pulsar or supernova remnant adds to the mystery as well, leaving one asking what is the possible emission mechanism at play in case of these unidentified UHE gamma-ray sources.

To that end, in this letter, we discuss and apply a phenomenological model, in which accelerated particles, escaped from an old, shell-type SNR (now invisible), interact with the associated MCs to produce the observed HE-VHE-UHE gamma-ray data for the case of LHAASO J2108+5157. We also provide the possible age of the old SNR, and account for the disappearance of the SNR in question. Our simple model is also consistent with the X-ray 2$\sigma$ upper limits given by \cite{abe22}. We also discuss the applicability of the model in other unidentified Galactic UHE gamma-ray source such as LHAASO J0341+5258. Furthermore, we report that the neutrino flux produced from the hadronic interaction considered in this model, will be non-detectable, even by the next generation observatory such as ICECUBE-Gen2 \citep{aartsen21}. 
   
\section{The model}\label{sec2}

In this section, we discuss the essentials of the model used to calculate the hadronic and leptonic components produced from the interaction between an old, now invisible SNR and the associated MCs. A more detailed discussion of the model can be found in \cite{desarkar22a}, where we developed and applied our model to explain the peculiar HE-VHE-UHE gamma-ray SED of LHAASO J1908+0621. Our simple model assumes that the supernova had exploded at the center of the cavity of a shell-like structure, which is surrounded by dense MCs. After this explosion, the SNR shock front expands inside the shell cavity, and finally hits the surrounding MCs. During the collision between the shock front and associated MCs, the accelerated particles get injected into the MCs to produce further emissions.

After the explosion, the supernova (SN) shock front expands freely during its free expansion phase. When the amount of swept-up interstellar medium (ISM) material becomes equal to that of the ejected material at t = t$_{Sedov}$, the SN enters the adiabatic Sedov phase. Finally after t = t$_{rad}$, the SN enters the radiative phase, in which the cooling timescales is less than the dynamic timescales. During its evolution, the time dependence of the shock velocity and radius can be given by the following simple relations \citep{fujita09, ohira12, desarkar22a},

\begin{equation}
\label{eq1}
\begin{split}
v_{sh}(t) = 
\begin{cases}
v_i & (t < t_{Sedov})\\
v_i(t/t_{Sedov})^{-3/5} & (t_{Sedov} < t)\\
\end{cases}              
\end{split}
\end{equation}

and, 

\begin{equation}
\label{eq2}
\begin{split}
R_{sh}(t) \propto 
\begin{cases}
(t/t_{Sedov}) & (t < t_{Sedov})\\
(t/t_{Sedov})^{2/5} & (t_{Sedov} < t)\\
\end{cases}              
\end{split}
\end{equation}

We note that for the entirety of this work, we have assumed the following values: initial velocity of the shock v$_i$ = 10$^9$ cm s$^{-1}$ \citep{fujita09}, radius of the shock and time at the beginning of the Sedov phase, R$_{Sedov}$ and t$_{Sedov}$, to be 2.1 pc and 210 yr, respectively \citep{ohira11, makino19}.

The CR protons are accelerated through Diffusive Shock Acceleration (DSA) mechanism when the SN is in the Sedov phase, where the CR protons accelerate by scattering back and forth across the shock front, while the shock is expanding towards the surrounding MCs. We assume an escape-limited acceleration scenario \citep{ohira10}, in which the CR protons need to escape a geometrical confinement region around the SN shock front produced by strong magnetic turbulence, in order to get injected into the MCs and take part in further interactions. The radius of the outermost boundary of this confinement region (i.e., the escape boundary) is called the escape radius, and it can be denoted by,

\begin{equation}
\label{eq3}
R_{esc}(t) = (1 + \kappa) R_{sh}(t),
\end{equation}

where $\kappa$ $\approx$ 0.04 \citep{ohira10,makino19}, and is defined by the geometrical confinement condition D$_{sh}$/v$_{sh}$ $\sim$ l$_{esc}$ = $\kappa$R$_{sh}$, where l$_{esc}$ is the distance of the escape boundary from the shock front and D$_{sh}$ is the diffusion coefficient around the shock \citep{ohira10}.

After the explosion, the escape boundary in front of the shock front eventually hits the surrounding MCs after traversing a distance of R$_{MC}$, the distance of MC surface from the cavity center. This essentially means that at the time of collision t$_{coll}$, the escape radius is equal to the MC surface distance, i.e. R$_{esc}$ (t$_{coll}$) = R$_{MC}$ $\approx$ R$_{sh}$ (t$_{coll}$), and at the time of collision, the velocity of the shock is denoted by v$_{sh}$(t$_{coll}$). We assume that the particle acceleration stops at t = t$_{coll}$ \citep{fujita09}. Consequently, protons accelerated at t $\leq$ t$_{coll}$ take part in further interactions inside the MCs. Moreover, only the protons with sufficiently high energies will escape the confinement region around the SNR shock front, whereas the low energy protons will remain confined around the SNR. So a suppression of fluxes in the lower energies, as well as a dominant contribution of fluxes in the highest energies should be expected in this scenario. The confinement condition invoked in this model changes the spectral shape of the injected proton population by constraining the minimum energy limit.

We estimate the minimum energy limit of the injected proton population by assuming that the escape
energy is a decreasing function of the shock radius \citep{makino19}. This approach is based on the assumption that the maximum energy of CR protons, E$_{max}^p$ is expected to increase up to knee energy ($\approx$ 10$^{15.5}$ eV) until the beginning of the Sedov phase, and then decrease from that epoch \citep{gabici09, ohira12}. The minimum energy required by protons to escape the confinement region can be given by the phenomenological relation, 

\begin{equation}
\label{eq4}
E_{esc}^p = E_{max}^p \left(\frac{R_{sh}}{R_{Sedov}}\right)^{-\alpha},   
\end{equation}

where $\alpha$ signifies the evolution of the maximum energy during the Sedov phase \citep{makino19}. We treat $\alpha$ as a free parameter in this work. After putting R$_{sh}$ $\approx$ R$_{esc}$ = R$_{MC}$ at the time of collision, we find the minimum energy required to escape the confinement zone, which also gives us the minimum energy threshold for the proton population that gets injected inside the surrounding MCs, i.e., E$_{esc}^p$ = E$_{min}^p$. Since protons are accelerated by DSA mechanism, we can expect the CR proton spectrum at the shock front $\propto$ E$^{-s}$. Then, in an escape-limited particle acceleration scenario, the protons with energies greater than $E_{esc}^p$ have a spectrum \citep{ohira10},

\begin{equation}
\label{eq5}
    N^p_{esc}(E) \propto E^{-[s + (\beta/\alpha)]},
\end{equation}

where $\beta$ represents a thermal leakage model of CR injection and is given by $\beta$ = 3(3–s)/2 \citep{makino19}.  For a typical value of s = 2, we get the value of $\beta$ = 1.5. Note that the spectral shape as well as the minimum energy of the proton population are calculated at the time when the escape boundary hits the surrounding MCs at t = t$_{coll}$.

At t $>$ t$_{coll}$, the shock enters the momentum conserving ``snowplow'' phase. The time evolution of the radius of the shocked shell R$_{shell}$ (t) inside the MCs can be found using momentum conservation equation \citep{fujita09, desarkar22a},

\begin{equation}
\label{eq6}
\begin{split}
\frac{4\pi}{3}&\left[n_{MC} ( R_{shell}(t)^3 -  R_{sh}(t_{coll})^3) + n_{cav}R_{sh}(t_{coll})^3\right]\Dot{R}_{shell}(t)\\
& = \frac{4\pi}{3}n_{cav}R_{sh}(t_{coll})^3 v_{sh}(t_{coll}), 
\end{split}
\end{equation}

with R$_{shell}$ = R$_{MC}$ at t = t$_{coll}$, n$_{MC}$ is the number density of the MCs, and n$_{cav}$ ($\approx$ 1 cm$^{-3}$) is the number density inside the cavity of the shell. We solve equation \ref{eq6} numerically for t $>$ t$_{coll}$, to estimate the current age of the SNR. We estimate the current age by considering the fact that the velocity of the shocked shell at the current age must be similar or even smaller than the internal gas velocity of the MCs. This approach takes into account the non-detection of any SNR shell in unidentified UHE gamma-ray sources discussed above, as the shell of the SNR becomes invisible owing to the higher internal gas velocity of the MCs as compared to that of the shocked shell. We consider the above discussed proton population and total number density of the cold protons inside the surrounding MCs (n$_{MC}$) to calculate total gamma-ray flux produced through hadronic interaction \citep{kafe}.

Similar to protons, electrons can also get accelerated in the SNR shock front and subsequently escape the confinement region to get injected in the associated MCs. Moreover, electrons also lose energy through radiative cooling very efficiently. Hence, the injected electron population was considered to be escape-limited, as well as loss-limited \citep{yamazaki06}. We consider the spectral index of the escaped electron population to be same as that of protons \citep{ohira12, desarkar22a}. To take into accout loss-limited nature of injected electron population, we consider a power law with exponential cutoff as the spectral shape of the escaped electrons,

\begin{equation}
\label{eq7}
    N^e_{esc}(E) \propto E^{-[s + (\beta/\alpha)]}exp(-E/E^e_{max}),
\end{equation}

where, maximum energy of the electron population has been determined by synchrotron cooling \citep{yamazaki06, fujita09},

\begin{equation}
     \label{eq8}
     E^e_{max} = 14h^{-1/2}\left(\frac{v_{sh}}{10^8\:\rm cm/s}\right)\left(\frac{B}{10\:\rm \mu G}\right)^{-1/2} \rm TeV,
\end{equation}

where, v$_{sh}$ is the velocity of the shock front and B is the downstream magnetic field. The parameter h (= $\frac{0.05r(f+rg)}{r-1}$, where r is the density compression ratio, f and g are functions of shock angle and gyro-factors) is used as a factor to calculate the acceleration time scale of DSA. We take h $\sim$ 1, considering the SNR in Sedov phase and neglecting non-linear effects, similar to \cite{yamazaki06}. We consider v$_{sh}$ = v$_{sh}$ (t$_{coll}$) since we calculate the maximum energy of the lepton population at the collision time and B = B$_{MC}$, the magnetic field inside the MCs. The minimum energy of the electron population was considered to be E$^e_{min}$ $\approx$ 500 MeV \citep{desarkar22a}. Furthermore, we consider bremsstrahlung, Inverse-Compton (IC) and synchrotron cooling \citep{blumenthal, ghisellini, baring} of the injected lepton population to calculate the gamma-ray flux produced. For IC interaction, we consider interstellar radiation field (ISRF) from \cite{popescu} at the source position, and the Cosmic Microwave Background (temperature T$_{CMB}$ = 2.7 K, energy density U$_{CMB}$ = 0.25 eV cm$^{-3}$) contribution as well. The number density was considered to be same as that of the MCs.

Finally we note that in this particular model, we have neglected the effect of diffusion of particles inside the MCs and assumed that the CR particles, both protons and electrons, lose energy through rapid cooling before escaping the cloud. This assumption can be realized by considering the idea that inside MCs, the diffusion is considerably suppressed (D $\approx$ 10$^{25-26}$ cm$^2$ s$^{-1}$) as compared to that observed in the ISM (D $\approx$ 10$^{28}$ cm$^{2}$ s$^{-1}$) \citep{gabici07, gabici09, fujita09, desarkar21}. Generation of plasma waves by CR streaming can be the reason behind the slow diffusion inside the MCs \citep{wentzel}. On the other hand, if the trapping of CR particles occurs due to some particular orientation of the magnetic field inside the MCs, then also the escape of the particles from the MCs will not be effective and can be neglected \citep{makino19}. 
Consequently, we have considered a steady-state proton and electron population to explain the SED of LHAASO J2108+5157, details of which are given in the next section.

\section{Application of the model: LHAASO J2108+5157}\label{sec3}

LHAASO J2108+5157 is an UHE gamma-ray source detected by LHAASO at R.A. = 317.22$^{\circ}$ $\pm$ 0.07$^{\circ}_{stat}$ and decl. = 51.95$^{\circ}$ $\pm$ 0.05$^{\circ}_{stat}$ \citep{cao2108} with a significance of 6.4$\sigma$ above 100 TeV. The source is reported to have a 95$\%$ confidence level extension upper limit of 0.26$^{\circ}$ with a 2D symmetrical Gaussian template, and its spectrum above 25 TeV can be well described by a power law with a photon index of 2.83 $\pm$ 0.18 \citep{cao2108}. Although no X-ray counterpart within 0.26$^{\circ}$ radius of the source was found, a spatially extended, HE counterpart 4FGL J2108.0+5155e (extension $\sim$ 0.48$^{\circ}$) \citep{abdollahi} was observed to be situated at an angular distance of 0.13$^{\circ}$ \citep{cao2108}. A new hard spectrum GeV source was also found at l = 92.35$^{\circ}$ and b = 2.56$^{\circ}$ by \textit{Fermi}-LAT data analysis \citep{abe22}, but its large angular separation ($\sim$ 0.27$^{\circ}$) from the LHAASO source indicates that this new source can hardly be a counterpart. Although no VHE component within 0.5${^\circ}$ radius was confirmed previously, recent observations by LST-CTA has hinted towards an existence of a source with 3.67$\sigma$ detection significance in the energy range of 3 - 100 TeV \citep{abe22}. Future observations may confirm an existence of a VHE counterpart with hard spectral index. The UHE source is located near the center of a GMC labeled [MML2017]4607 \citep{miville}, which has an average angular radius and mass of 0.236$^{\circ}$ and 8469 M$_{\odot}$, respectively, and is situated at a distance of 3.28 kpc from Earth. The average number density of the GMC was estimated to be n$_{MC}$ $\approx$ 30 cm$^{-3}$ \citep{cao2108}. The presence of the GMC, spatially coincident with the UHE gamma-ray source points towards the hadronic origin, but leptonic origin can not be neglected. The absence of any energetic pulsar, its wind nebula or SNR warrants a cautious approach in unveiling the true nature of emission regarding this UHE source.  

Two young open stellar clusters Kronberger 80 and Kronberger 82 are in the vicinity of the LHAASO source (with angular distances of 0.62$^{\circ}$ and 0.45$^{\circ}$, respectively) \citep{cao2108}. But large angular separation between these clusters and LHAASO source centroid, as well as absence of proper distance estimation hint that the contribution of these clusters are unlikely \citep{cao2108, abe22}. \cite{cao2108} suggested that UHE gamma-ray emission is due to an interaction of escaping CRs with MCs, whereas the GeV counterpart maybe due to an old SNR. However, \cite{abe22} pointed out that photon index of GeV counterpart spectrum is too soft compared to the observations of old SNRs
interacting with MCs \citep{yuan12}, and to produce UHE gamma-ray spectrum, the required spectral index of the proton population has to be very hard as compared to the standard DSA theory. Instead, \cite{abe22} proposed an alternate leptonic scenario, in which UHE gamma-ray emission is due to TeV halo emission, and the GeV counterpart is due to a tentative, previously undetected pulsar. But a very low associated magnetic field (even lower than the average Galactic magnetic field), and non-detection of a pulsar make the TeV halo interpretation questionable, and open the source up for further exploration. To that end, we apply the model discussed in Section \ref{sec2} to explain the gamma-ray data from HE to UHE energy range, while being consistent with the X-ray 2$\sigma$ upper limits. We note that these 2$\sigma$ X-ray upper limits correspond to a uniform, circular source with a radius of 6$'$ centered on the position of the LHAASO source \citep{abe22}. We explain the VHE-UHE gamma-ray data with hadronic component produced from the interaction between protons, accelerated and escaped at an early time from a now old SNR shock front, with protons inside the surrounding MCs, whereas the HE gamma-ray data is explained using bremsstrahlung cooling of accelerated and escaped electrons inside the medium of the MCs. Our model also shows that the main contribution in X-ray range comes from the synchrotron cooling of the same accelerated and escaped electrons.

\begin{figure}
\includegraphics[width=\columnwidth]{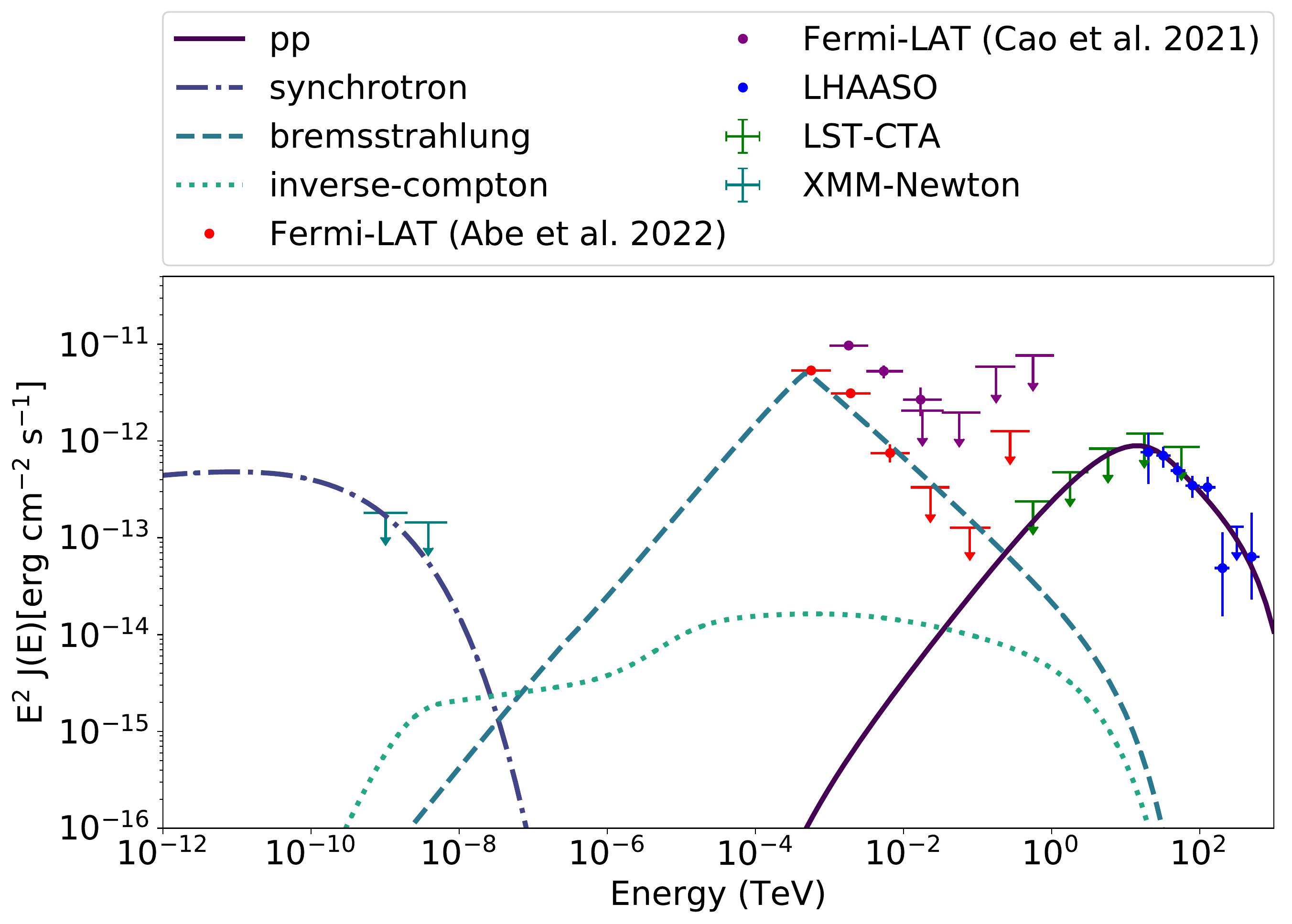}
\caption{MWL SED of LHAASO J2108+5157. Gamma-ray data points and upper limits obtained from different observatories such as \textit{Fermi}-LAT (red \citep{abe22}, purple \citep{cao2108}), LHAASO (blue \citep{cao2108}), and LST-CTA (green \citep{abe22}) are shown in the figure. The XMM-Newton X-ray 2$\sigma$ upper limits \citep{abe22} are given in teal. The model p-p interaction (solid line), bremsstrahlung (dashed), IC (dotted), and synchrotron (dot-dashed) components are also shown in the figure.}
\label{fig1}
\end{figure}

\begin{figure}
\includegraphics[width=\columnwidth]{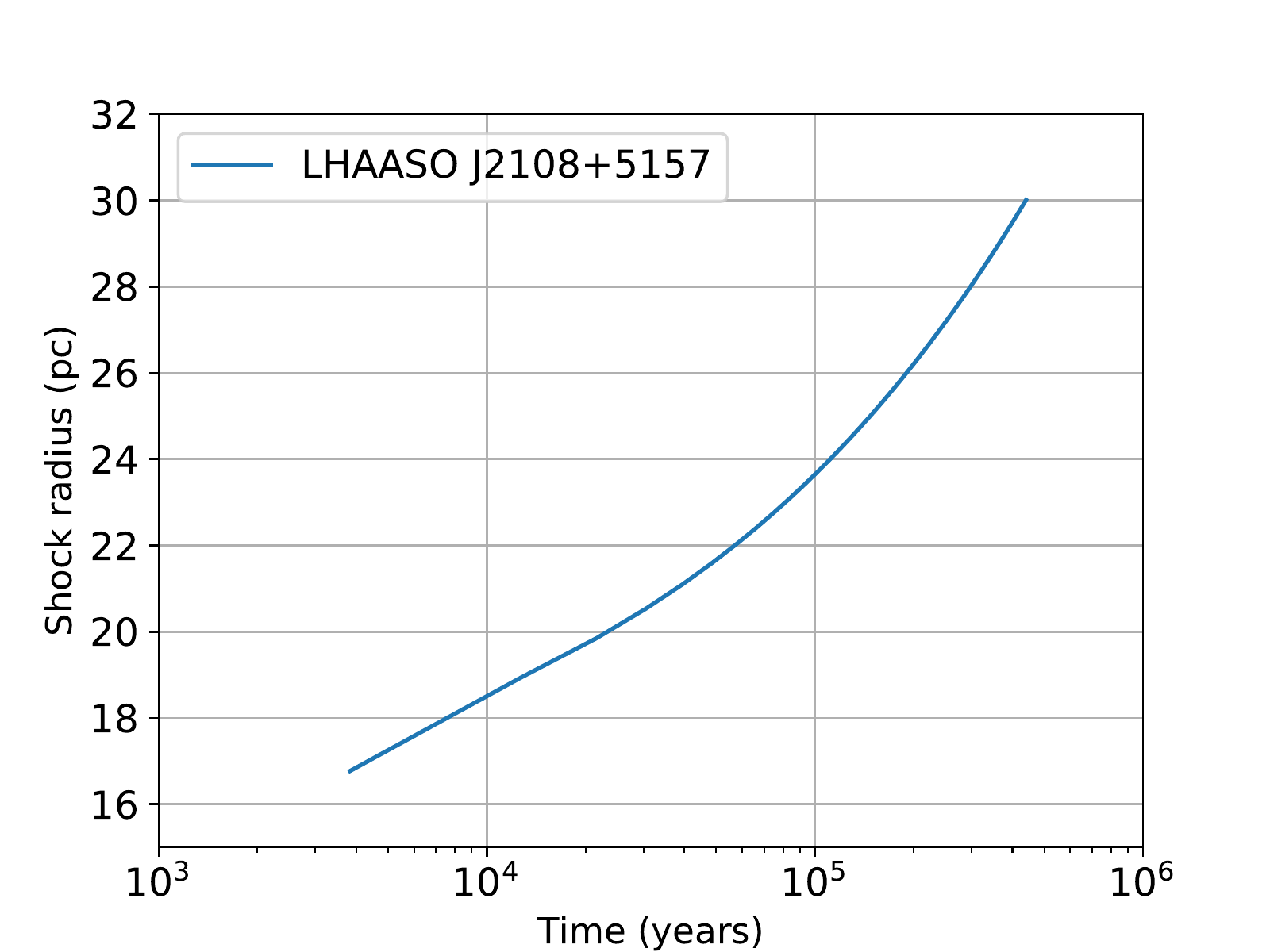}
\caption{Time evolution of the shocked shell associated with the old SNR, inside the surrounding MCs.}
\label{fig2}
\end{figure}

In this work, we have considered the free parameter $\alpha$ = 1.875, and then let the total energy budgets of proton and electron populations vary to explain the MWL SED. Considering the value of $\alpha$, and the values of s and $\beta$ discussed in Section \ref{sec2}, we get the spectral indices of the escaped electron and proton populations as p = [s + ($\beta$/$\alpha$)] = 2.8. The distance of the source was taken to be d $\sim$ 3 kpc. The model spectrum components, as well as the considered MWL SED are shown in Figure \ref{fig1}. Also, we calculate the time evolution of SNR shocked shell inside the associated MCs using equation \ref{eq6}, and find that the SNR, with a final radius of $\sim$ 30 pc, has to be $\sim$ 4.4 $\times$ 10$^5$ years old, for the shock velocity to be lower than the internal gas velocity of MC [MML2017]4607 ($\sim$ 13 km s$^{-1}$) \citep{cao2108}, and the SNR shell to disappear. The time evolution of the shocked shell is shown in Figure \ref{fig2}. Finally, the model parameters required to explain the gamma-ray data are shown in Table \ref{tab1}. We have used open source code \texttt{GAMERA} \citep{hahn} to calculate the model spectrum of different components.

\begin{table}
 \caption{Parameters Used in The Model.}
 \label{tab1}
 \begin{tabular*}{\columnwidth}{l@{\hspace*{5pt}}l@{\hspace*{5pt}}l}
  \hline
  Definition & Parameter & Value\\
  \hline
  \textit{SNR/MC structure and evolution:}\\
  \hline
  Initial shock velocity & v$_i$ (cm/s) & 10$^9$\\ 
  Time at the start of Sedov phase & t$_{Sedov}$ (years) & 210\\
  Shock radius at the start of Sedov phase & R$_{Sedov}$ (pc) & 2.1\\
  Time of collision & t$_{coll}$ (years) & 3.83 $\times$ 10$^3$\\
  Shock radius at time of collision & R$_{sh}$ (t$_{coll}$) (pc) & 16.77 (= R$_{MC}$)\\
  Shock velocity at time of collision & v$_{sh}$ (t$_{coll}$) (cm/s) & 1.75 $\times$ 10$^8$\\
  Current age of SNR & t$_{age}$ (years) & 4.4 $\times$ 10$^5$\\
  Final radius of shock & R$_{sh}$ (t$_{age}$) (pc) & 30\\
  Final velocity of shock & v$_{sh}$ (t$_{age}$) (cm/s) & 1.2 $\times$ 10$^6$\\ 
  Distance & d (kpc) & 3\\
  MC number density & n$_{MC}$ (cm$^{-3}$) & 30\\
  MC magnetic field & B$_{MC}$ ($\mu$G) & 25\\
  Cavity number density & n$_{cav}$ (cm$^{-3}$) & 1\\
  \hline
  \textit{Hadronic component:}\\
  \hline
  Minimum energy & E$^p_{min}$ (TeV) & 63\\
  Maximum energy & E$^p_{max}$ (TeV) & 3.1 $\times$ 10$^3$\\
  Spectral index & p & 2.8\\
  Energy budget & W$_p$ (erg) & 3.6 $\times$ 10$^{47}$\\
  \hline
  \textit{Leptonic component:}\\
  \hline
  Minimum energy & E$^e_{min}$ (TeV) & 5 $\times$ 10$^{-4}$\\
  Maximum energy & E$^e_{max}$ (TeV) & 15.5\\
  Spectral index & p & 2.8\\
  Energy budget & W$_e$ (erg) & 3.6 $\times$ 10$^{47}$\\
  \hline
 \end{tabular*}
\end{table}

\section{Discussion and Conclusion}\label{sec4}

In this letter, we have discussed and applied a simple, analytical and phenomenological model to explain the HE-VHE-UHE gamma-ray data observed from the direction of LHAASO J2108+5157. By only adjusting the index $\alpha$, not only we show that the model components are consistent with gamma-ray and X-ray observations, the results also naturally explain the observed morphology of the source region, e.g., the disappearance of the SNR at current age. As expected, the SNR was found be old ($>$ 10$^5$ years). This also explains why no pulsar has been seen in the source region, as the pulsar is expected to leave the source region due to its natal kick velocity ($\sim$ 400-500 km s$^{-1}$) \citep{gaensler06}. Similar nature and emission were also found in another UHE gamma-ray source, LHAASO J1908+0621, details of which were explained by this model in \cite{desarkar22a}. The fact that the emission of multiple UHE gamma-ray sources were explained by the same model hints towards its validity in a larger context. Interestingly, another unidentified UHE gamma-ray source, LHAASO J0341+5258, also shows similar characteristics shown by LHAASO J2108+5157 \citep{cao0341}. It is very likely that this model is applicable in that case as well. However, in that case, the VHE counterpart has not been properly constrained, and the High Altitude Water Cherenkov (HAWC) upper limit provided in \cite{cao0341} corresponds to only a 2$\sigma$ detection significance. Further observations by CTA and detailed analysis by \textit{Fermi}-LAT will be necessary to properly constrain the emission of LHAASO J0341+5258. 

From Figure \ref{fig1}, we can see that the hadronic component adequately explain the VHE-UHE gamma-ray data, whereas the bremsstrahlung component, originated from the cooling of the electron population, explains the gamma-ray data in the HE range. The bremsstrahlung component is expected to dominate the IC component, as the interaction is taking place inside MCs with a high number density of cold protons. Moreover, the synchrotron component does not violate the X-ray 2$\sigma$ upper limits. We note that no proper radio counterpart has been associated with the LHAASO J2108+5157 yet. An extended radio source associated with nearby star-forming region \citep{cao2108}, as well as point-like radio source NVSS 210803+515255 or WENSS B2106.4+5140 \citep{abe22} were found within 95$\%$ extension upper limit of LHAASO J2108+5157 and 4FGL J2108.0+5155e. Since no proper association was established between these sources and the gamma-ray source, we refrain from including their radio data in this study to further constrain the model, and we follow the MWL SED discussed in \citep{abe22} to ascertain the feasibility of the model discussed in this letter.  

As discussed earlier, we have neglected the effect of particle diffusion in this model. We note that such an assumption may likely lead to an overestimation, and the aspect of suppressed diffusion inside the MCs is highly uncertain \citep{xu16,dogiel15}. Introducing an energy-independent diffusion coefficient, as discussed in \cite{dogiel15}, will lead to higher energy budgets required by the electron and proton populations to explain the data. The suppressed diffusion coefficient introduced by \cite{gabici09} has similar energy dependence as to that observed in ISM, but the exact energy dependence of diffusion coefficient inside clouds is not well constrained. So, to avoid further complications, we have neglected the effect of diffusion in this model, similar to \cite{fujita09, makino19}, and assumed that the injected particles quickly cool down before escaping the MC medium.

We further note that we do not consider the contribution of accelerated and escaped particles, when the shock front is within the MC medium, in calculating the total gamma-ray SED. Even if the SNR is still in the Sedov phase when the shock is within the MCs, the corresponding contribution was found to be negligible. Moreover, the acceleration and subsequent escape of particles, in that case,  will depend on the evolution of the confinement region within the high-density, turbulent medium of the MCs, details of which is beyond the scope of the simple model discussed in this letter. Furthermore, as the SNR enters its radiative phase at t$_{rad}$ $\sim$ 4 $\times$ 10$^4$ years \citep{blondin98}, the particle acceleration becomes ineffective as the small shock velocity at that age, as obtained from equation \ref{eq6} ($<$ 1.1 $\times$ 10$^7$ cm/s), prevents full ionization of the pre-shock gas \citep{shull79}. So no significant contribution to the total gamma-ray SED is expected in the radiative phase of the SNR as well.

Since hadronic component primarily dominates in the VHE-UHE gamma-ray range, neutrinos can be produced from the hadronic interaction as well. This neutrino flux can be a smoking gun evidence for the dominant hadronic interaction. We have calculated the neutrino flux resulting from the hadronic interaction discussed above, and found that the corresponding neutrino flux is too low to be detected by current generation neutrino telescope such as ICECUBE. Furthermore, we have found that the model neutrino flux does not exceed the 5$\sigma$ discovery potential after 10 years of observation by next generation neutrino observatory ICECUBE-Gen2 for two declinations, $\delta$ = 0$^{\circ}$ and 30$^{\circ}$ \citep{aartsen21}, which indicates that it is unlikely to confirm the hadronic nature of UHE gamma-ray emission through neutrino observations, even in the near future, for this source.

In conclusion, in this letter, we have shown that by essentially tuning the $\alpha$ index, the emission of the LHAASO source can be explained. We note that we do not intend to ``fit'' the MWL SED, as the SED, in various energy ranges (VHE, X-ray, radio), is poorly constrained and in need of further observations. In this work, we have only applied a simple phenomenological model, while also minimizing the free parameters, which naturally explains the spectral features and spatial morphology of LHAASO J2108+5157. Future observations can confirm the viability of this model to explain LHAASO J2108+5157, or other unidentified UHE gamma-ray source LHAASO 0341+5258, and sources detected in future as well, which show similar nature and emission signatures. If confirmed, then it can be posited that SNRs as a source class, similar to PWNe, can likely be a strong candidate for being the Galactic PeVatrons.   

\section*{Acknowledgements}

I thank the anonymous reviewer for helpful comments and constructive criticism. I thank Nayantara Gupta for encouragement.

\section*{Data Availability}

The simulated data underlying this paper will be shared on reasonable request to the corresponding author.
 


\bibliographystyle{mnras}
\bibliography{example} 








\bsp	
\label{lastpage}
\end{document}